\documentclass[sn-mathphys,Numbered]{sn-jnl}

\usepackage{graphicx}%
\usepackage{multirow}%
\usepackage{amsmath,amssymb,amsfonts}%
\usepackage{amsthm}%
\usepackage{mathrsfs}%
\usepackage[title]{appendix}%
\usepackage{xcolor}%
\usepackage{textcomp}%
\usepackage{manyfoot}%
\usepackage{booktabs}%
\usepackage{algorithm}%
\usepackage{algorithmicx}%
\usepackage{algpseudocode}%
\usepackage{listings}%

\theoremstyle{thmstyleone}%
\theoremstyle{thmstyletwo}%

\theoremstyle{thmstylethree}%

\begin{document}

\title{Topologically enhanced nonlinear optical response of graphene nanoribbon heterojunctions}


\author*[1]{\fnm{Hanying} \sur{Deng}}\email{dhy0805@alumni.sjtu.edu.cn}
\author[1]{\fnm{Zhihao} \sur{Qu}}
\author[1]{\fnm{Yingji} \sur{He}}  
\author[2]{\fnm{Changming} \sur{ Huang}}
\author*[3]{\fnm{Nicolae C.} \sur{Panoiu}}\email{n.panoiu@ucl.ac.uk}
\author*[4]{\fnm{ Fangwei} \sur{Ye}}\email{fangweiye@sjtu.edu.cn}

\affil*[1]{\orgdiv{School of Optoelectronic Engineering}, \orgname{Guangdong Polytechnic Normal University}, \orgaddress{\city{Guangzhou}, \postcode{510665}, \country{China}}}

\affil[2]{\orgdiv{Department of Physics}, \orgname{Changzhi University}, \orgaddress{ \city{Shanxi}, \postcode{046011}, \country{China}}}

\affil*[3]{\orgdiv{Department of Electronic and Electrical Engineering}, \orgname{University College London}, \orgaddress{\street{Torrington Place}, \city{London}, \postcode{WC1E 7JE}, \country{United Kingdom}}}

\affil*[4]{\orgdiv{School of Physics and Astronomy}, \orgname{Shanghai Jiao Tong University}, \orgaddress{\city{Shanghai}, \postcode{200240}, \country{China}}}


\abstract{We study the nonlinear optical properties of heterojunctions made of graphene nanoribbons (GNRs) consisting of two segments with either the same or different topological properties. By utilizing a quantum mechanical approach that incorporates distant-neighbor interactions, we demonstrate that the presence of topological interface states significantly enhances the second- and third-order nonlinear optical response of GNR heterojunctions that are created by merging two topologically inequivalent GNRs. Specifically, GNR heterojunctions with topological interface states display third-order harmonic hyperpolarizabilities that are more than two orders of magnitude larger than those of their similarly sized counterparts without topological interface states, whereas the second-order harmonic hyperpolarizabilities exhibit a more than ten-fold contrast between heterojunctions with and without topological interface states. Additionally, we find that the topological state at the interface between two topologically distinct GNRs can induce a noticeable red-shift of the quantum plasmon frequency of the heterojunctions. Our results reveal a general and profound connection between the existence of topological states and an enhanced nonlinear optical response of graphene nanostructures and possible other photonic systems.}

\maketitle

\section{Introduction}\label{sec1}
The topology of matter has become a subject of extensive research interest due to its ability to explain many intriguing physical phenomena, such as topological insulators \cite{RevModPhys.83.1057,RevModPhys.82.3045}, the quantum spin Hall
effect \cite{PhysRevLett.95.146802, PhysRevLett.95.226801}, and Majorana fermions \cite{fermino5} and their analogs. In particular, interfaces between two materials of different topological classes give rise to highly robust topological states \cite{konig2007quantum,PhysRevLett7,wang2017topological}. Since the first discovery of topological states in condensed matter physics, these states have been observed in a variety of systems spanning different areas of science, including mechanics \cite{PhysRevLett9,huber2016topological,PhysRevB11}, acoustics \cite{he2016acoustic,ye2022topological,PhysRevLett14}, and optics \cite{lu2014topologicalcheck,Topologica16,2016Topological,RevModPhy18,deng2019topological}. Recently, localized topological states have been achieved at the interface between two segments of graphene nanoribbons (GNRs) with different topological phases \cite{PhysRevLettc20,2018Topological21,PhysRevLett22}. The topological phases of GNRs are determined by their width, edge, and end termination \cite{PhysRevLettc20,2018Topological21}. With recent advances in bottom-up synthesis from precursor molecules, atomically precise GNRs with various widths, well-defined edges, and end terminations can be readily produced \cite{cai2014graphene23,ruffieux2012electronic24,liu2020bottom25,narita2015bottom}, providing a rich variety of graphene systems suitable for further development of possible applications.

Despite the extensive study and rapid progress in topological physics, most of existing studies have focused on the linear regime. However, recent research suggests that the interplay between topology and nonlinearity can give rise to important novel phenomena \cite{mittal2018topological27,you2020four,wang2019topologically29,smirnova2019third}. For instance, a topological source of quantum light has been realized in a two-dimensional array of ring resonators \cite{mittal2018topological27}, whereas four-wave mixing of topological edge plasmons has been observed in graphene metasurfaces \cite{you2020four}. In addition, strong third-harmonic generation has been experimentally demonstrated in a SSH-like left-handed transmission line \cite{wang2019topologically29}. Moreover, GNRs inherit the intrinsically large nonlinearity from extended graphene sheets and produce large field enhancement \textit{via} plasmon excitation \cite{cox2016quantum,cox2019nonlinear,hendry2010coherent33}. Consequently, GNR heterojunctions composed of two topologically inequivalent GNRs provide an ideal and versatile platform to investigate the interplay between topology and optical nonlinearity at the nanoscale.

When the size of graphene nanostructures is less than $\sim$ 10 nm, the widely used classical nonlinear conductivity is inadequate to describe their nonlinear optical properties. This is so because nonlocal and finite-size effects play a crucial role in determining their optical response, necessitating full-quantum calculations \cite{cox2014electrically34,manjavacas2013plasmons35,manrique2017quantum36,deng2018quantum37,deng2021quantum}. One widely used quantum mechanical method to describe the electronic states of graphene nanostructures is the tight-binding  model \cite{cox2015plasmon,ezawa2007metallic}. However, the standard tight-binding model considers only the coupling of nearest-neighbor atoms, so that the accuracy of its predictions can be improved. Recently, we proposed a distant-neighbor quantum mechanical (DNQM) approach to compute the linear and nonlinear optical properties of graphene nanostructures \cite{deng2018quantum37,deng2021quantum}. Compared to the tight-binding model, the DNQM calculations are more accurate, as it includes the coupling of the $\pi$-orbital electron of each atom with the core potential of all atoms in the graphene nanostructure. In particular, our DNQM method has been applied to calculate the second- and third-order nonlinear optical response of certain graphene nanostructures \cite{deng2021quantum}.

In this work, we use DNQM calculations to investigate the nonlinear optical properties of GNR heterojunctions composed of two segments of GNRs with either the same or different topological phases. We show that due to the emergence of topological interface states, the nonlinear optical response of GNR heterojunctions consisting of two topologically distinct GNRs is significantly enhanced. More specifically, compared to GNR heterojunctions without topological interface states, the third-order nonlinear hyperpolarizability of GNR heterojunctions with topological interface states can increase by more than two orders of magnitude. Additionally, second-harmonic generation in GNR heterojunctions with topological interface states is shown to be more than 10 times larger than that in topologically trivial heterojunctions with similar size. We also find that the resonance frequency of quantum plasmons in GNR heterojunctions is red-shifted when the topological interface states are generated.

\section{System configuration}

\begin{figure}[ht!]
\centering\includegraphics[width=10cm]{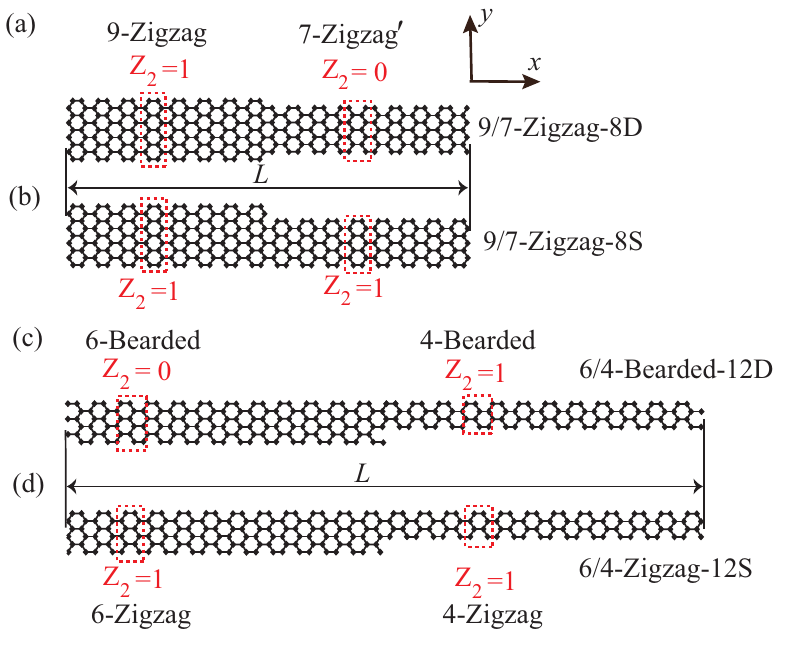}
\caption{Armchair GNR heterojunctions formed with two GNRs of (a) W=9 zigzag and $W=7$ zigzag$^{\prime}$ terminations, (b) $W=9$ zigzag and $W=7$ zigzag terminations, (c) $W=6$ and $W=4$ bearded terminations, and (d) $W=6$ zigzag and $W=4$ zigzag terminations. $W$ is the number of rows of carbon atoms along the lateral direction. The two GNR segments in (a) and (c) are topologically inequivalent, whereas those in (b) and (d) are topologically equivalent. The dashed red rectangles indicates a unit cell.}
\label{fig1}
\end{figure}

The GNR heterojunctions we considered are shown in Fig.~\ref{fig1}. They consist of two armchair-edged GNRs ($W$-AGNRs) that may or may not be topologically equivalent. Here, $W$ represents the number of rows of carbon atoms that form the width of GNRs. The topology of GNRs is determined by their width, edge, and end termination, and can be characterized by a $Z_2$ invariant~\cite{PhysRevLettc20,2018Topological21}. This invariant is an integer and takes a value of $Z_2=1$ ($Z_2=0$) for topologically nontrivial (trivial) ribbons with a unit cell that possesses spatial inversion and/or mirror symmetry. We first focus on the AGNRs with zigzag, zigzag$^{\prime}$, and bearded terminations, as they are the most commonly synthesized forms of GNRs. For odd $W$, the values of $Z_2$ of $W$-AGNRs are given by \cite{PhysRevLettc20}:
\begin{subequations}\label{Z2odd}
\begin{align}\label{eq1}
&Z_2=\frac{1+(-1)^{[\frac{W}{3}]+[\frac{W+1}{2}]}}{2}\quad \mathrm{zigzag-terminated},\\
\label{eq2}
&Z_2=\frac{1-(-1)^{[\frac{W}{3}]+[\frac{W+1}{2}]}}{2}\quad \mathrm{zigzag^{\prime}-terminated}.
\end{align}
\end{subequations}
Here, $[x]$ is the floor function that takes the maximum integer smaller than or equal to the real number $x$. For even $W$, the values of $Z_2$ of AGNRs are \cite{PhysRevLettc20}:
\begin{subequations}\label{Z2even}
\begin{align}\label{eq3}
&Z_2=\frac{1-(-1)^{[\frac{W}{3}]+[\frac{W+1}{2}]}}{2}\quad \mathrm{zigzag-terminated},\\
\label{eq4}
&Z_2=\frac{1-(-1)^{[\frac{W}{3}]}}{2}\quad \mathrm{bearded-terminated}.
\end{align}
\end{subequations}

Knowing the value of the $Z_2$ invariant of $W$-AGNRs with various terminations and widths, we construct four heterojunctions lying on the $x$-$y$ plane, at which two AGNRs of same or distinct topology are joined together,
as shown in Fig.~\ref{fig1}.

The two heterojunctions presented in Figs.~\ref{fig1}(a) and \ref{fig1}(b) consist of a zigzag-terminated $W=9$ AGNR with $Z_2=1$ and a zigzag$^{\prime}$-terminationed $W=7$ AGNR with $Z_2=0$ [Fig.~\ref{fig1}(a)], and a zigzag-terminated $W=7$ AGNR with $Z_2=1$ [Fig.~\ref{fig1}(b)]; we labeled them as 9/7-Zigzag-8D and 9/7-Zigzag-8S, respectively. Note that the two heterojunctions contain the same number of carbon atoms, $N_c=256$, and have the same length of $L=6.7$ \text{nm}. We also consider two heterojunctions containing the same number of carbon atoms, $N_c=240$, and having the same side length of $L=10.1$ \text{nm} by connecting two bearded- or zigzag-terminated AGNRs with even rows of carbon atoms, $W=4$ and 6, and denote them as 6/4-Bearded-12D and 6/4-Zigzag-12S, as shown in Figs.~\ref{fig1}(c) and 1(d), respectively. The two AGNR segments in Fig.~\ref{fig1}(c) are topologically inequivalent, while those in Fig.~\ref{fig1}(d) are topologically equivalent.

\section{Energy spectra and charge density distributions}
\begin{figure}[ht!]
\centering\includegraphics[width=10cm]{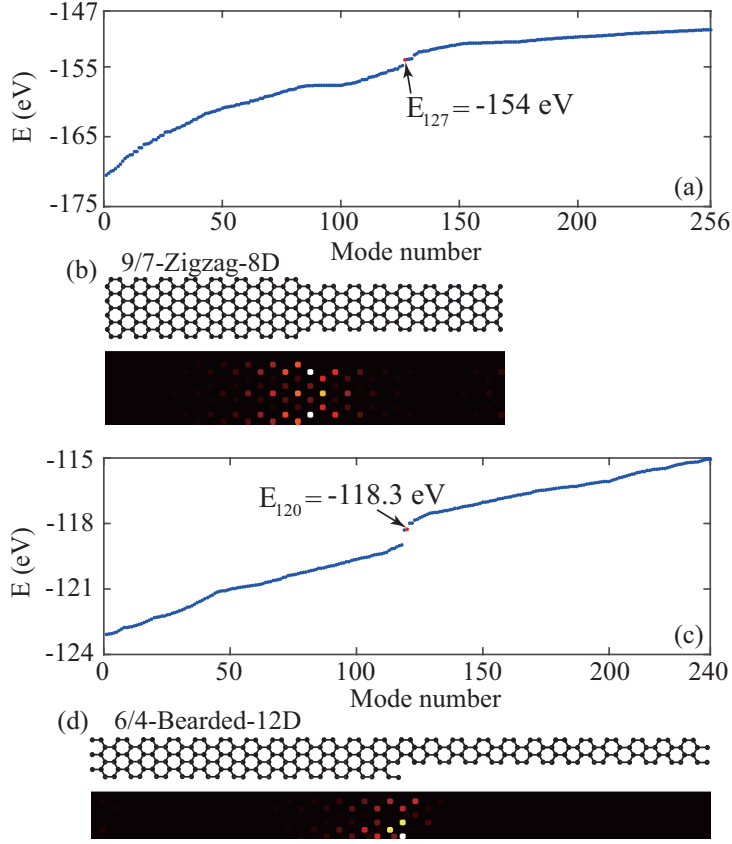}
\caption{Energy spectra of (a) 9/7-Zigzag-8D and (c) 6/4-Bearded-12D heterojunctions. A topological interface state exists at the interface of (b) 9/7-Zigzag-8D and (d) 6/4-Bearded-12D heterojunctions. The topological interface states in 9/7-Zigzag-8D and 6/4-Bearded-12D heterojunctions are marked by the red dots in (a) and (c), respectively.}
\label{fig2}
\end{figure}

We first investigate energy spectra and charge density distributions of the AGNR heterojunctions using the DNQM method (details about the DNQM calculations can be found in Appendix A). Figures~\ref{fig2}(a) and~\ref{fig2}(c) depict the energy spectra of the 9/7-Zigzag-8D and 6/4-Bearded-12D heterojunctions, respectively, with modes arranged in increasing order of their enenergy. We also calculate their corresponding charge density distributions. In accordance with the bulk-edge correspondence principle, our computations demonstrate that a topologically induced interface state arises at the 9/7-Zigzag-8D heterojunction, localized at the interface between the two topologically distinct AGNR segments, as shown in Fig.~\ref{fig2}(b). Remarkably, the energy of the topological interface state in the 9/7-Zigzag-8D heterojunction corresponds to the HOMO-1 (with HOMO meaning the highest-occupied molecular orbital), having a value of $E=-154$ eV. The red dot in the energy spectrum in Fig.~\ref{fig2}(a) indicates this energy level.

Similarly, Fig.~\ref{fig2}(d) illustrates a localized interface state of the 6/4-Bearded-12D heterojunction, due to the fact that the two GNR segments possess different topological phases. We also note that the energy of the topological interface state of the 6/4-Bearded-12D heterojunction corresponds to the HOMO, with a value of -118.3 eV, as indicated by the red dot in the energy spectrum in Fig.~\ref{fig2}(c). By contrast, no localized interface states exist at GNR heterojunctions formed with two topologically equivalent GNR segments (9/7-Zigzag-8S and 6/4-Zigzag-12S).

\section{Linear and nonlinear optical response}

We next investigate the linear and nonlinear optical response of the AGNR heterojunctions composed of two GNRs that are topologically equivalent and inequivalent. We first consider the 9/7-Zigzag-8D and 9/7-Zigzag-8S heterojunctions. As discussed above, the 9/7-Zigzag-8D heterojunction has a topological state located at the interface separating its two topologically inequivalent AGNR segments, whereas no localized interface states exist in the 9/7-Zigzag-8S heterojunction. Additionally, the two AGNR heterojunctions have the same length and contain the same number of carbon atoms. Notice that, because its symmetry to reflection w.r.t. the $x-z$ plane, the second-order nonlinear optical response of the 9/7-Zigzag-8D heterojunction is not allowed for $y$-polarized incident fields. Thus, we assume that the incident electric field is $x$-polarized. The results of the calculations for the linear polarizability and second- and third-order hyperpolarizabilities of 9/7-Zigzag-8D and 9/7-Zigzag-8S heterojunctions are summarized in Fig.~\ref{fig3} (see calculation details in Appendix B).

\begin{figure}[ht!]
\centering\includegraphics[width=10cm]{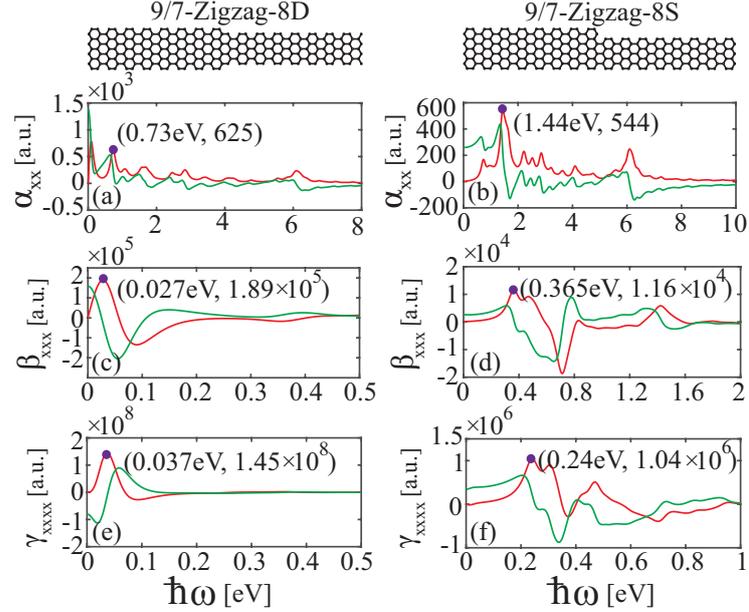}
\caption{Real and imaginary parts of [(a), (b)] linear polarizability, [(c), (d)] second-order, and [(e), (f)] third-order hyperpolarizabilities of 9/7-Zigzag-8D and 9/7-Zigzag-8S heterojunctions, respectively. The red (green) curves indicate the imaginary (real) part of the polarizabilities. The purple dots denote the resonance peaks. The polarizabilities are given in atomic units (a.u.) with $e=\hbar=m_{e}=a_{0}=1$.}
\label{fig3}
\end{figure}

In Figs.~\ref{fig3}(a) and~\ref{fig3}(b), the real and imaginary parts of the linear polarizability spectra, $\alpha_{xx}(\omega)$, of 9/7-Zigzag-8D and 9/7-Zigzag-8S heterojunctions are shown, respectively. Quantum plasmons exist when the peak of the imaginary parts of the polarizability occurs at a zero of the real 
part\cite{manrique2017quantum36,deng2018quantum37, 2013Quantum}. From Figs.~\ref{fig3}(a) and \ref{fig3}(b), we can see that the first resonance frequency of quantum plasmon in the topologically nontrivial heterojunction is substantially smaller than that in the topologically trivial heterojunction. Specifically, the 9/7-Zigzag-8D and 9/7-Zigzag-8S heterojunctions have the first resonance peak located at 0.73 eV and 1.44 eV, respectively. This feature can be explained by the molecular orbital associated with its topological interface state participating in the transition. Thus, the first resonance peak in the linear polarizabilities of the 9/7-Zigzag-8D heterojunction is located at ${\hbar}\omega=0.73$ eV, which corresponds to the transition from the HOMO-1 state (which as discussed above is the topological state) to the LUMO+2.

Figures~\ref{fig3}(c) and~\ref{fig3}(d) show the real and imaginary parts of the second-order hyperpolarizability, $\beta_{xxx}(2\omega)$, corresponding to the second-harmonic generation (SHG), of 9/7-Zigzag-8D and 9/7-Zigzag-8S heterojunctions, respectively. Remarkably, the SHG hyperpolarizability spectra of these two heterojunctions are significantly different. Firstly, it can be seen that $\beta_{xxx}(2\omega)$ of topologically nontrivial 9/7-Zigzag-8D heterojunction is more than one order of magnitude larger than that of topologically trivial 9/7-Zigzag-8S heterojunction. Additionally, similar to the linear case, the resonance frequency of the second-order nonlinear plasmon of 9/7-Zigzag-8D heterojunction is red-shifted in comparison to that of 9/7-Zigzag-8S heterojunction. These two interesting characteristics of the topologically nontrivial 9/7-Zigzag-8D heterojunction can be ascribed to its topological interface state participating in the transitions. In particular, as shown in Fig.~\ref{fig3}(c), the most pronounced peak in the spectrum of SHG hyperpolarizabilities of 9/7-Zigzag-8D heterojunction is located at ${\hbar}\omega=0.027$ eV, which corresponds to the transition from the HOMO-1, corresponding to the topological interface state, to the LUMO, indicating that the presence of the topological interface state can dramatically enhance the second-order nonlinear response of the 9/7-Zigzag-8D heterojunction and induce a red-shift of the resonance frequency of its quantum plasmons.

In Figs.~\ref{fig3}(e) and~\ref{fig3}(f), we show the spectra of the third-harmonic generation (THG) hyperpolarizability, $\gamma$$_{xxxx}(3\omega)$, of 9/7-Zigzag-8D and 9/7-Zigzag-8S heterojunctions, respectively. Similar to the case of the second-order nonlinear optical response, due to the presence of the topological interface state, the THG of the topologically nontrivial 9/7-Zigzag-8D heterojunction is dramatically enhanced and its plasmon frequency is red-shifted. Interestingly, we find that the hyperpolarizability $\gamma$$_{xxxx}(3\omega)$ of the topologically nontrivial 9/7-Zigzag-8D heterojunction is enhanced by more than two orders of magnitude in comparison to that of the topologically trivial 9/7-Zigzag-8S heterojunction.

\begin{figure}[ht!]
\centering\includegraphics[width=10cm]{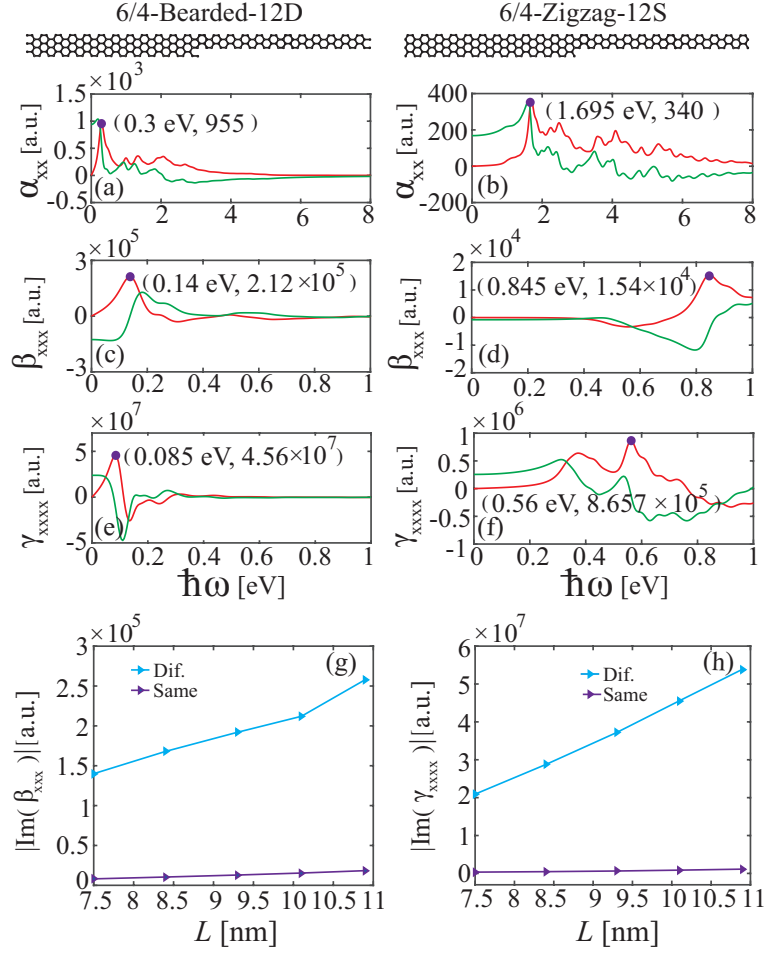}
\caption{Real and imaginary parts of [(a), (b)] linear,  [(c), (d)] second-order, and [(e),(f)] third-order hyperpolarizability of 6/4-Bearded-12D and 6/4-Zigzag-12S heterojunctions. [(g), (h)] Evolution of the peak value of (g) SHG and (h) THG hyperpolarizabilities with the side length of 6/4-Bearded-12D (blue curves) and 6/4-Zigzag-12S (purple curves) heterojunctions, respectively. The red (green) curves indicate the imaginary (real) part of the hyperpolarizabilities. The purple dots denote the resonance peaks.}
\label{fig4}
\end{figure}

We now consider the 6/4-Bearded-12D and 6/4-Zigzag-12S heterojunctions and use the DNQM approach to investigate their linear and nonlinear optical response, the corresponding results being summarized in Fig.~\ref{fig4}. As explained above, the 6/4-Bearded-12D heterojunction possesses a topological interface state, while no topological interface states exist in the 6/4-Zigzag-12S heterojunction. The real and imaginary parts of linear polarizability, $\alpha_{xx}(\omega)$, of 6/4-Bearded-12D and 6/4-Zigzag-12S heterojunctions are presented in Figs.~\ref{fig4}(a) and~\ref{fig4}(b), respectively. Again, due to the emergence of the topological interface state, the most pronounced peak in the spectrum of $\alpha_{xx}(\omega)$ of 6/4-Bearded-12D heterojunction is significantly red-shifted in comparison to that of 6/4-Zigzag-12S heterojunction. The nonlinear hyperpolarizabilities associated with SHG ($\beta_{xxx}(2\omega)$) and THG ($\gamma$$_{xxxx}(3\omega)$) are presented in Figs.~\ref{fig4}(c,d) and~\ref{fig4}(e,f), respectively, for the 6/4-Bearded-12D and 6/4-Zigzag-12S heterojunctions. Comparing these spectra, we can draw similar conclusions to those revealed by Fig.~\ref{fig3}. More specifically, due to the existence of the topological interface state, both the SHG and THG hyperpolarizabilities of the 6/4-Bearded-12D heterojunction exceed by more than one order of magnitude those of the topologically trivial 6/4-Zigzag-12S heterojunction. Furthermore, the presence of topological interface states also leads to an obvious red-shift of the nonlinear plasmon frequency of 6/4-Bearded-12D heterojunction.

In Figs.~\ref{fig4}(g) and~\ref{fig4}(h), we present an overview of the dependence of the maximum SHG and THG hyperpolarizabilities on the side length of the 6/4-Bearded-12D and 6/4-Zigzag-12S heterojunctions, respectively. We can see that the SHG and THG hyperpolarizabilities of 6/4-Bearded-12D heterojunction that possesses a topological interface state are more than one order of magnitude larger than those of the topologically trivial 6/4-Zigzag-12S heterojunction of similar side length. As expected, the magnitudes of the SHG and THG hyperpolarizabilities of both 6/4-Bearded-12D and 6/4-Zigzag-12S heterojunctions increase with the increase of their side length. Interestingly enough, we note that the rate of the increase of the nonlinear hyperpolarizabilities with side length for the topologically nontrivial 6/4-Bearded-12D heterojunction is larger than that for the topologically trivial 6/4-Zigzag-12S heterojunction.

\begin{figure}[ht!]
\centering\includegraphics[width=10cm]{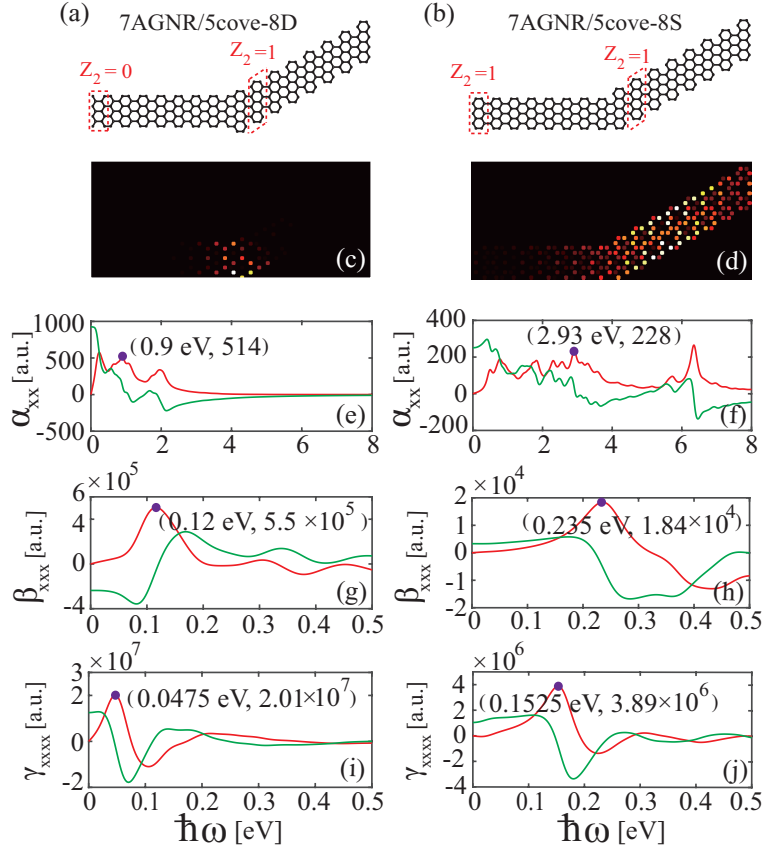}
\caption{GNR heterojunctions formed with (a) a zigzag$^{\prime}$-terminationed $W=7$ AGNR and a $P=5$ cove-edged GNR and (b) a zigzag-terminated $W=7$ AGNR and a $P=5$ cove-edged GNR. The two GNR segments in (a) and (b) are topologically inequivalent and equivalent, respectively. Charge density distributions of (c) 7AGNR/5cove-8D and (d) 7AGNR/5cove-8S. Real and imaginary parts of [(e), (f)] linear polarizability and [(g), (h)] second-order, and [(i), (j)] third-order  hyperpolarizability of 7AGNR/5cove-8D and 7AGNR/5cove-8S heterojunctions. The red (green) curves indicate the imaginary (real) part of the polarizabilities. The purple dots denote the resonance peaks.}
\label{fig5}
\end{figure}

Finally, we focus on a more complex GNR heterojunction composed of an AGNR and a cove-edged GNR, as schematically shown in Figs.~\ref{fig5}(a) and \ref{fig5}(b). Cove-edged GNRs have been recently synthesized by bottom-up process from precursor molecules \cite{liu2015toward}. Their topological properties are determined by the ribbon width, edge shape, and end termination \cite{2018Topological21}. Figures~\ref{fig5}(a) and \ref{fig5}(b) present two heterojunctions consisting of a $P=5$ cove-edged GNR with tilting angle of $30^{\text{o}}$ ($Z_2=1$) and either a zigzag$^{\prime}$-terminationed $W=7$ AGNR ($Z_2=0$) or a zigzag-terminated $W=7$ AGNR ($Z_2=1$), which are denoted by 7AGNR/5cove-8D and 7AGNR/5cove-8S, respectively. Here, $P$ denotes the number of zigzag chains forming the width of the cove-edged GNR. The two heterojunctions contain the same number of carbon atoms, $N_c=256$. Again, by bulk-boundary correspondence principle, a topological interface state occurs for the 7AGNR/5cove-8D heterojunction and is localized at the interface between the two GNRs of distinct topology, as shown in Fig.~\ref{fig5}(c). By contrast, for the 7AGNR/5cove-8S heterojunction comprised of two topologically equivalent GNRs, no localized interface states exist and the corresponding charge profile is extended throughout the heterojunction, as shown in Fig.~\ref{fig5}(d).

We show in  Figs.~\ref{fig5}(e-j) the frequency dependence of linear polarizability, and SHG and THG hyperpolarizabilities of the 7AGNR/5cove-8D and 7AGNR/5cove-8S heterojunctions, respectively. As expected, due to the occurrence of the topological interface state, the resonance frequency of both the linear and nonlinear plasmons of 7AGNR/5cove-8D heterojunction is significantly red-shifted, as compared with that of 7AGNR/5cove-8S heterojunction. Furthermore, the magnitudes of the second- and third-order nonlinear hyperpolarizabilities of the topologically nontrivial 7AGNR/5cove-8D heterojunction are significantly larger than those of the topologically trivial 7AGNR/5cove-8D heterojunction, as indicted by the red dots in Figs.~\ref{fig5}(g-j), respectively.

\section{Conclusion}
In conclusion, we have utilized the DNQM approach to investigate the optical response of GNR heterojunctions, comprising of two topologically equivalent or inequivalent ribbons. Our findings reveal a significant enhancement in the nonlinear optical response of GNR heterojunctions that possess topological interface states as compared to the trivial ones. Specifically, we have observed an increase in the third-order hyperpolarizabilities of GNR heterojunctions with topological interface states by over two orders of magnitude in comparison to those without such states, despite having the same number of carbon atoms. Moreover, due to the presence of topological interface states, the second-order hyperpolarizabilities of GNR heterojunctions with topologically distinct GNRs are over one order of magnitude larger than those of topologically trivial heterojunctions. Furthermore, we have noticed a large red-shift in the resonance frequency of quantum plasmons in GNR heterojunctions that contain topological states. Remarkably, these results were established for several heterojunctions with different configurations, which suggests that there is a general and deep connection between the existence of topological states and enhanced nonlinear optical response in GNRs and other photonic systems.

\begin{appendices}


\section*{Appendix A: Calculation of energy spectra and charge density distributions of GNR heterojunctions}

We use a DNQM approach to calculate the energy spectra and charge density distributions of GNR heterojunctions shown in Fig.1 and Fig.5. Our calculations include the interactions between $\pi$ electrons and all cores of carbon atoms in the GNR heterojunctions. It goes beyond the tight-binding approximation used successfully to determine the linear and nonlinear optical response of graphene nanoflakes~\cite{cox2016quantum, cox2019nonlinear, cox2014electrically34}, as in our case the coupling constants are calculated from first principles. Specifically, we assume that  the $\pi$ electrons reside in the $2p_z$ carbon orbitals oriented perpendicular to the GNR heterojunction plane and populate on average with one such electron per carbon site. Thus, we employ the Clementi orbital~\cite{Clementi} as a basis function on each atom, which takes the form of

\begin{align}
\psi_{2p_{z}}=R_{2p}(r)Y_{2p_{z}}(\theta,\varphi)\tag{A1},
\end{align}
with
\begin{align}
R_{2p}(r)=\frac{1}{\sqrt{6}} \frac{Zr}{na_0}Z^{\frac{3}{2}}e^{-\frac{Zr}{na_0}}\tag{A2},
\end{align}
and 
\begin{align}
Y_{2p_{z}}(\theta, \varphi)=\sqrt{\frac{3}{4\pi}}\cos\theta =\sqrt{\frac{3}{4\pi}}\frac{z}{r}\tag{A3}.
\end{align}
Here, $a_0$ is the Bohr radius, $n=2$ is the orbital number, and $Z=3.136$ is the effective nuclear charge for the $2p_z$ orbital of a carbon atom\cite{Clementi}.

The Hamiltonian operator for a single electron in  a GNR heterojunction is 
\begin{align}
\widehat{H}=-\frac{\hbar}{2m}\nabla_{\vec{r}}^{2}-\sum_{\alpha=1}^N\frac{Z_{\text{eff}}e^2}{\vec{r}-\vec{r}_{0\alpha}}\tag{A4},
\end{align}
where the value of effective core charge, $Z_{\text{eff}}=0.637$, has been adjusted so that the computed
HOMO-LUMO gap of the simplest GNR structure (i.e. benzene) is 6 eV~\cite{ezawa2007metallic}.

The energy spectra and charge density distributions are calculated using the the Schr\"{o}dinger equation

\begin{align}
\widehat{H}\psi(\vec{r})= E\psi(\vec{r})\tag{A5},
\label{eqa5}
\end{align}
where $E$ and $\psi(\vec{r})$ are the eigenenergy and eigenfunction, respectively. This
eigenfunction is expanded in the atom-centered basis functions as
\begin{align}
\psi(\vec{r})=\sum_{q=1}^Nc_q\psi_{q}(\vec{r})=\sum_{q=1}^Nc_q\psi_{2p_z}(\vec{r}-\vec{r}_{oq})\tag{A6}.
\label{eqa6}
\end{align}
From Eq.~(\ref{eqa5}) and Eq.~(\ref{eqa6}), we can have
\begin{align}
\sum_{q=1}^N\widehat{H}c_q\psi_{q}=E\sum_{q=1}^Nc_q\psi_{q}\tag{A7}.
\label{eqa61}
\end{align}

We then use $\psi_{s}^{*}(\vec{r})$, $s=1,...,N$, to multiply Eq.~(\ref{eqa61}), and integrate
over the entire space:
\begin{align}
\sum_{q=1}^Nc_q\int \psi_{s}^*\widehat{H}\psi_{q}d\vec{r}=E\sum_{q=1}^Nc_q \int
\psi_{s}^*\psi_{q}d\vec{r}\tag{A8}.
\end{align}
Now, we define the matrix elements
\begin{subequations}
\begin{align}
\widehat{H}_{sq}&=\int\psi_{s}^*\hat{H}\psi_{q}d\vec{r}\tag{A9a}, \\
\widehat{S}_{sq}&=\int\psi_{s}^*\psi_{q}d\vec{r}\tag{A9b},
\end{align}
\end{subequations}
so that the Schr\"{o}dinger equation is reduced to a finite generalized eigenvalue problem with
eigenvectors, $\widehat{c}$, and eigenvalues, $E$
\begin{align}
\widehat{H}\hat{c}=E\widehat{S}\hat{c}\tag{A10},
\label{eq:10}
\end{align}
where $\hat{c}=(c_{1},c_{2},\ldots,c_{N})^{T}$ is an $N$-dimensional column vector. We can obtain the energy spectra and charge density distributions of a GNR heterojunction containing $N$ carbon atoms by numerically solving Eq.~(\ref{eq:10}).

\section*{Appendix B: Quantum perturbative approach to the linear and nonlinear optical polarizabilities of GNR heterojunctions}

We calculate linear and nonlinear optical polarizabilities of GNR heterojunctions using a well-known quantum perturbative approach~\cite{Boyd}.
The linear polarizability, $\alpha_{ij}(\omega)$, and the nonlinear hyperpolarizabilities corresponding to second-harmonic generation, $\beta_{ijk}(2\omega)$, and third-harmonic generation,
$\gamma_{hijk}(3\omega$) are given by~\cite{Boyd}

\begin{align}
\alpha_{ij}(\omega)=\sum_{g,m}^{transition}\Big(
\frac{\mu_{gm}^{i}\mu_{mg}^{j}}{\omega_{mg}-\omega}+\frac{\mu_{gm}^{j}\mu_{mg}^{i}}{\omega_{mg}^{*}+\omega}
\Big)\tag{B1},
\label{eqa19}
\end{align}

\begin{align}
\beta_{ijk}(2\omega)=P_l\sum_{g,m,n}^{transition}\Big[\frac{\mu_{gn}^{i}\mu_{nm}^{j}\mu_{mg}^{k}}{(\omega_{ng}-2\omega)(\omega_{mg}-\omega)}
&+\frac{\mu_{gn}^{j}\mu_{nm}^{i}\mu_{mg}^{k}}{(\omega_{ng}^{*}+\omega)(\omega_{mg}-\omega)}\nonumber \\
&+\frac{\mu_{gn}^{j}\mu_{nm}^{k}\mu_{mg}^{i}}{(\omega_{ng}^{*}+\omega)(\omega_{mg}^{*}+2\omega)}\Big]\tag{B2},
\label{eqa20}
\end{align}

\begin{align}
\gamma_{hijk}((3\omega)=&P_l\sum_{g,m,n,v}^{transition}\Big[\frac{\mu_{gv}^{h}\mu_{vn}^{i}\mu_{nm}^{j}\mu_{mg}^{k}}{(\omega_{vg}-3\omega)(\omega_{ng}-2\omega)(\omega_{mg}-\omega)} \nonumber \\
 & +\frac{\mu_{gv}^{i}\mu_{vn}^{h}\mu_{nm}^{j}\mu_{mg}^{k}}{(\omega_{vg}^{*}+\omega)(\omega_{ng}-2\omega)(\omega_{mg}-\omega)} 
  +\frac{\mu_{gv}^{i}\mu_{vn}^{j}\mu_{nm}^{h}\mu_{mg}^{k}}{(\omega_{vg}^{*}+\omega)(\omega_{ng}^{*}+2\omega)(\omega_{mg}-\omega)}\nonumber \\
  &+\frac{\mu_{gv}^{i}\mu_{vn}^{j}\mu_{nm}^{k}\mu_{mg}^{h}}{(\omega_{vg}^{*}+\omega)(\omega_{ng}^{*}+2\omega)(\omega_{mg}^{*}+3\omega)}\Big]\tag{B3},
\label{eqa21}
\end{align}
where  $\vec{\mu}=e\vec{r}$ is the dipole moment operator, $\mu_{gm}=\int\psi_{g}^*\vec{\mu}\psi_{m}d\vec{r}$ is the transition dipole moment, $\omega_{mg}=\frac{E_m-E_g}{\hbar}-i\eta$, ${\psi_m}=\sum_{l}c_{ml}{\psi_l}$, and
$E_m$ and $c_m=\sum_{l}c_{ml}$ are the eigenenergy and engenvector of the eigenstate $m$, respectively. In addition, $g$, $m$, $n$ and $v$ are labels used to distinguish between levels of transitions, and $\eta= 0.1$ eV is
related to the lifetime of excited states. $P_l$ is a intrinsic permutation operator that defines the expression to be summed over all permutations of the Cartesian indices: $h$, $i$, $j$, and $k$.




\end{appendices}

\backmatter


\end{document}